\documentclass[useAMS,usenatbib]{mn2e}
\usepackage{aas_macros,graphicx,times,multirow,amsmath}
\usepackage[]{color}
\title[Observations of \src\ and PSR\,B1830--08]
  {Long-term spectral and timing properties of the soft gamma-ray repeater SGR\,1833--0832 and detection of extended X-ray emission around the radio pulsar PSR\,B1830--08}
\author[P. Esposito et al.]
{P.~Esposito,$^{1}$\thanks{E-mail: paoloesp@oa-cagliari.inaf.it} G.~L.~Israel,$^{2}$ R.~Turolla,$^{3,4}$ F.~Mattana,$^{5}$ A.~Tiengo,$^{6}$  A. Possenti,$^{1}$ 
\newauthor  S.~Zane,$^{4}$ N.~Rea,$^{7}$ M. Burgay,$^{1}$ D.~G\"otz,$^{8}$ S.~Mereghetti,$^{6}$ L.~Stella,$^{2}$ M.~H.~Wieringa,$^{9}$ 
\newauthor J.~M.~Sarkissian,$^{9}$ T.~Enoto,$^{10}$ P. Romano,$^{11}$ T.~Sakamoto,$^{12}$ Y.~E.~Nakagawa,$^{13}$ 
\newauthor K.~Makishima,$^{13,14}$ K.~Nakazawa,$^{14}$ H.~Nishioka,$^{14}$ C.~Fran\c{c}ois-Martin$^{15}$
\smallskip\\
$^1$INAF -- Osservatorio Astronomico di Cagliari, localit\`a Poggio dei Pini, strada 54, I-09012 Capoterra, Italy\\
$^2$INAF -- Osservatorio Astronomico di Roma, via Frascati 33, I-00040 Monteporzio Catone, Italy\\
$^3$Universit\`a di Padova, Dipartimento di Fisica, via F.~Marzolo 8, I-35131 Padova, Italy\\
$^4$Mullard Space Science Laboratory, University College London, Holmbury St. Mary, Dorking, Surrey RH5 6NT, UK\\
$^5$Fran\c{c}ois Arago Centre, APC (Universit\'e Paris Diderot, CNRS/IN2P3, CEA/DSM, Observatoire de Paris), 13 rue Watt, F-75205 Paris Cedex 13, France\\
$^6$INAF -- Istituto di Astrofisica Spaziale e Fisica Cosmica - Milano, via E.~Bassini 15, I-20133 Milano, Italy\\
$^7$Institut de Ci\`encies de l'Espai (CSIC--IEEC), Campus UAB, Facultat de Ci\`encies, Torre C5-parell, E-08193 Barcelona, Spain\\
$^8$AIM (UMR 7158 CEA/DSM-CNRS-Universit\'e Paris Diderot) Irfu/Service d'Astrophysique, Saclay,  F-91191
Gif-sur-Yvette Cedex, France\\
$^{9}$Australia Telescope National Facility, CSIRO Astronomy and
Space Science, P.O. Box 76, Epping 1710, Australia\\
$^{10}$Kavli Institute for Particle Astrophysics \& Cosmology (KIPAC), SLAC/Stanford University, PO Box 20450, MS 29, Stanford, CA 94309, USA\\
$^{11}$INAF -- Istituto di Astrofisica Spaziale e Fisica Cosmica - Palermo, via U.~La~Malfa 153, I-90146 Palermo, Italy\\
$^{12}$NASA Goddard Space Flight Center, Greenbelt, MD 20771, USA\\
$^{13}$High Energy Astrophysics Laboratory, Institute of Physical and Chemical Research (RIKEN), Wako, Saitama, 351-0198, Japan\\
$^{14}$Department of Physics, University of Tokyo, 7-3-1 Hongo, Bunkyo-ku, Tokyo, 113-0033, Japan\\
$^{15}$Universit\'e Denis Diderot - Paris 7, 4 place Jussieu, F-75252, Paris Cedex 5, France}
\date{Accepted 2011 May 6.  Received 2011 May 4; in original form 2011 April 12}
\pagerange{\pageref{firstpage}--\pageref{lastpage}} \pubyear{2011}

\def\LaTeX{L\kern-.36em\raise.3ex\hbox{a}\kern-.15em
    T\kern-.1667em\lower.7ex\hbox{E}\kern-.125emX}
    
\def\xmm {\emph{XMM-Newton}}
\def\cxo {\emph{Chandra}}
\def\swift {\emph{Swift}}

\def\rxte {\emph{RXTE}}
\def\rst {\emph{ROSAT}}

\def\src {SGR\,1833--0832}
\def\flux {\mbox{erg cm$^{-2}$ s$^{-1}$}}
\def\lum {\mbox{erg s$^{-1}$}}
\def\nh {$N_{\rm H}$}

\begin{document}

\label{firstpage}
\maketitle
\begin{abstract}
\src\ was discovered on 2010 March 19 thanks to the \swift\ detection of a short hard X-ray burst and follow-up X-ray observations. Since then, it was repeatedly observed with \swift, \emph{Rossi X-ray Timing Explorer}, and \xmm. Using these data, which span about 225 days, we studied the long-term spectral and timing characteristics of \src. We found evidence for diffuse emission surrounding \src, which is most likely a halo produced by the scattering of the point source X-ray radiation by dust along the line of sight, and we show that the source X-ray spectrum is well described by an absorbed blackbody, with temperature $kT\sim1.2$ keV and absorbing column $N_{\rm H}=(10.4\pm0.2)\times10^{22}$ cm$^{-2}$, while different or more complex models are disfavoured. The source persistent X-ray emission remained fairly constant at $\sim$$3.7\times10^{-12}$ \flux\ for the first $\sim$20 days after the onset of the bursting episode, then it faded by a factor $\sim$40 in the subsequent $\sim$140 days, following a power-law trend with  index $\alpha\simeq-0.5$. We obtained a phase-coherent timing solution with the longest baseline ($\sim$225 days) to date for this source which, besides period $P=7.565\,408\,4(4)$ s and period derivative $\dot{P}=3.5(3)\times10^{-12}$ s s$^{-1}$, includes higher order period derivatives.  We also report on our search of the counterpart to the SGR at radio frequencies using the Australia Telescope Compact Array and the Parkes radio telescope. No evidence for  radio emission was found, down to flux densities of 0.9 mJy (at 1.5 GHz) and 0.09 mJy (at 1.4 GHz) for the continuum and pulsed emissions, respectively, consistently with other observations at different epochs. Finally, the analysis of the field of  PSR\,B1830--08 (J1833--0827), which was serendipitously  imaged by the \xmm\ observations, led to the discovery of the X-ray pulsar wind nebula generated by this 85-ms radio pulsar. We discuss its possible association with the unidentified TeV source HESS\,J1834--087.

\end{abstract}

\section{Introduction}

Soft gamma-ray repeaters (SGRs; seven confirmed members) and anomalous X-ray pulsars (AXPs; twelve confirmed members)\footnote{See the McGill Pulsar Group catalogue at the webpage\\ \mbox{http://www.physics.mcgill.ca/$\sim$pulsar/magnetar/main.html}.} are two classes of X-ray pulsating sources with no evidence for companion stars which share a number of properties. These include rotation periods of several seconds ($P\sim2$--$12$ s), rapid spin down ($\dot{P}\sim10^{-11}$ s s$^{-1}$), large and variable X-ray luminosities (exceeding the rate of rotational energy loss), and the emission of flares and short bursts (see \citealt{woods06,mereghetti08} for reviews). SGRs and AXPs (note that the distinction is becoming increasingly blurred) are currently interpreted as observational manifestations of magnetars, namely neutron stars powered by their huge magnetic field (e.g. \citealt{paczynski92,duncan92,thompson95,thompson96}; \citealt*{tlk02}). This picture is supported by the fact that the dipole magnetic fields inferred for SGRs and AXPs from their period and period derivative\footnote{According to the usual magnetic braking formula, for a neutron star of 10 km radius and 1.4 solar masses, $B_{\rm dip}\approx3.2\times10^{19}(P\dot{P})^{1/2}$ G (e.g. \citealt{lorimer04}).} are above, or at the high end of, those of the radio pulsars. Surface magnetic fields in SGRs/AXPs in fact often exceed $10^{14}$ G, although an upper limit as low as $7.5\times10^{12}$ G has been recently reported for SGR 0418+5729 \citep{rea10short}.

Between periods of activity, characterised by bursts and significant variability in flux, spectrum, pulse shape, and spin-down rate,  magnetars go through long stretches of quiescence. The discovery of the first `transient' AXP, XTE\,J1810--197 \citep{ibrahim04short},\footnote{Actually, at that time another magnetar with a transient behaviour, SGR\,1627--41 \citep{kouveliotou03short,eiz08short}, was already known, but its characteristics were not well established yet. Significant flux variability was reported also in the AXP candidate AX\,J1845.0--0300 \citep{torii98,vasisht00,tam06}.} showed that during quiescence magnetars can be faint and not dissimilar from hundreds of unidentified sources present in various X-ray catalogues (such as the \rst, \xmm, and \cxo\ ones). This suggested that a potentially large number of magnetars had not been discovered yet and may manifest themselves in the future. Thanks also to the effectiveness of the \swift\ and \emph{Fermi} satellites in catching magnetar outbursts, five new magnetars (all of them transients) were discovered in the last few years and several major outbursts were observed from known sources (see \citealt{rea11} and references therein). 

A recent addition to the magnetar family is \src. It was discovered on 2010 March 19 when,  at 18:34:50 UT, the \swift\ Burst Alert Telescope (BAT) triggered on a short ($<$1 s) hard X-ray burst and localised it in a region close to the Galactic plane \citep{gelbord10short,barthelmy10short,gogus10short}. The \swift\ X-ray Telescope (XRT) started observing the BAT field one minute after the trigger and unveiled the existence of a previously unknown bright X-ray source. Given the proximity to the Galactic plane and the burst properties, the X-ray source was immediately suggested to be an SGR. The SGR nature of the source, now catalogued as \src, has been confirmed shortly after by the discovery with \swift\ and \emph{Rossi X-ray Timing Explorer} (\rxte) of pulsations at 7.57 s \citep{gogus10atel,eis10,palmer10}. Subsequent observations also allowed the determination of the spin-down rate of \src\ \citep{eib10short,gogus10short}. Following the onset of the outburst, the source flux remained fairly constant for about 20 days \citep{gogus10short}. No radio, optical or infrared counterparts have been detected \citep{burgay10short,wieringa10short,gogus10short}.
 
Here we report on the spatial and long-term spectral and temporal behaviour of the persistent X-ray emission of \src\ using new \xmm, \rxte\ and \swift\ observations. In Sections~\ref{observations} and \ref{results} we describe the X-ray observations used in our study and we present the results of our analysis. In Section~\ref{radio-obs} we give more details on the radio observations presented in \citet{burgay10short} and \citet{wieringa10short}. In Section~\ref{hess} we report on the analysis of the field of  PSR\,B1830--08 (J1833--0827), which was serendipitously  imaged by the \xmm\ observations, and discuss its possible association with the unidentified TeV source HESS\,J1834--087. Discussion follows in Section~\ref{disc}.

\section{X-ray observations}\label{observations}

\subsection{\xmm}
\begin{table}
\centering
\caption{Journal of the \swift\ and \xmm\ observations. The time of the BAT trigger is MJD 55274.774.}
\label{obs-log}
\begin{tabular}{@{}lccc}
\hline
Instrument & Obs.ID & Start date & Exposure\\
 & & (MJD) & (ks)\\
\hline
\swift & 00416485000 & 55274.775 & 29.1\\
\swift & 00416485001 & 55276.177 & 10.8\\
\swift & \phantom{$^{a}$}00416485002$^{a}$ & 55276.700 & 10.0\\
\swift & 00416485003 & 55277.327 & 9.9\\
\xmm   & 0605851901 & 55278.525 & 22.0\\
\swift & 00416485004 & 55278.374 & 8.1\\
\swift & 00416485005 & 55279.047 & 10.3\\
\swift & 00416485006 & 55280.067 & 9.8\\
\swift & 00416485007 & 55281.520 & 10.0\\
\swift & 00416485008 & 55282.457 & 9.9\\
\swift & 00416485009 & 55283.003 & 10.9\\
\swift & 00416485010 & 55284.531 & 9.5\\
\swift & 00416485011 & 55286.014 & 7.9\\
 \xmm  & 0605852001 & 55288.503 & 21.0\\
\swift & 00416485012 & 55289.625 & 10.0\\
\swift & 00416485013 & 55293.573 & 10.1\\
\swift & 00416485014 & 55298.406 & 5.1\\
\swift & 50041648015 & 55299.133 & 4.0\\
\xmm   & 0605852101 & 55299.187 & 19.0\\
\swift & 00416485016 & 55301.003 & 9.4\\
\swift & 00416485017 & 55304.623 & 8.8\\
\swift & 00416485018 & 55307.163 & 10.3\\
\swift & 00416485019 & 55309.117 & 7.6\\
\swift & 00416485020 & 55315.653 & 5.5\\
\swift & 00416485021 & 55316.055 & 4.4\\
\swift & 00416485022 & 55339.813 & 18.0\\
\swift & 00416485023 & 55432.162 & 5.3\\
\swift & 00416485024 & 55433.372 & 2.2\\
\swift & 00416485025 & 55434.106 & 9.9\\
\swift & 00416485026 & 55435.242 & 2.5\\
\hline
\end{tabular}
\begin{list}{}{}
\item[$^{a}$] This observation was carried out in WT mode.
\end{list}
\end{table}

The three focal plane CCD cameras of the \xmm\ EPIC instrument, pn \citep{struder01short}, MOS1 and MOS2 \citep{turner01short}, cover the 0.1--12 keV energy range with an effective area of roughly 1400 cm$^2$ for the pn and 600 cm$^2$ for each MOS. After the discovery of \src, \xmm\ pointed its mirrors towards the new SGR three times (see Table~\ref{obs-log}). All observations were performed with the thick optical filter and in full frame mode,\footnote{See the \xmm\ User Handbook at\\ http://xmm.esac.esa.int/external/xmm\_user\_support/documentation/index.shtml.} except for the first one, carried out with the MOS cameras in large window mode. Moreover, \xmm\ serendipitously imaged the field of \src\ on 2006 September 16 during an observation targeting the nearby supernova remnant G23.5--0.0 (obs. ID: 0400910101, exposure: 12.4 ks; all the detectors were in full frame mode with the medium filter).

The data were processed using version 10.0 of the \xmm\ Science Analysis Software (\textsc{sas}) and standard screening criteria were applied. Source events were accumulated for each camera from circular regions with a 36 arcsec radius. We selected this aperture, corresponding to $\sim$85\% of the encircled energy fraction at 5 keV for a point source, in order to minimise the contamination from the diffuse emission surrounding \src\ (see Section~\ref{diffuse}). The background counts were extracted from source-free regions far from the position of the SGR. The ancillary response files and the spectral
redistribution matrices for the spectral analysis were generated with the \textsc{sas} tasks \textsc{arfgen} and \textsc{rmfgen}, respectively.

\subsection{\swift}

The X-Ray Telescope (XRT; \citealt{burrows05short}) on-board \swift\ uses a CCD detector sensitive to photons between 0.2 and 10 keV with an effective area of about 110 cm$^2$. Twenty-seven observations of \src\ were performed (see Table~\ref{obs-log}), starting right after its discovery, in both photon counting (PC) and windowed timing (WT) modes (see \citealt{hill04short} for more details on the XRT readout modes). The results of the first 18 \swift\ observations have already been published in \citet{gogus10short}, while the others are reported here for the first time.\\
\indent The data were processed and filtered with standard criteria using the \textsc{ftools} software package. We extracted the PC source events from a circle with a radius of 15 pixels (one pixel corresponds to about $2\farcs36$) and the WT data from a 25-pixel-wide strip. To estimate the background, we extracted PC and WT events from source-free regions distant from the position of \src. For the spectral fitting we used the latest available spectral redistribution matrix in \textsc{caldb} (v011), while the ancillary response files were generated with \textsc{xrtmkarf}, and they account for different extraction regions, vignetting and point-spread function corrections. 

\subsection{\rxte}

The Proportional Counter Array (PCA; \citealt{jahoda96}) on-board \rxte\ consists of an array of five collimated xenon/methane multi-anode Proportional Counter Units (PCUs) operating in the 2--60 keV energy range, with a total effective area of approximately 6500 cm$^2$ and a full width at half-maximum field of view (FOV) of about 1$\degr$. The 80 \rxte/PCA pointed observations of \src\ (obs. ID: 95048) reported here span from 2010 March 19 to 2010 December 04. The exposures range from 1 ks to 20 ks, for a total of about 466.4 ks.  The results of the first 31 \rxte\ observations have already been published in \citet{gogus10short}, while the analysis of the others is reported here for the first time.

The raw data were reduced using the \textsc{ftools} package. Given the non-imaging nature and wide FOV of the PCA instrument and the relatively low flux of \src, the \rxte\ data were used only to study the timing properties of the source. We thus restricted our analysis to the data in Good Xenon mode, with a time resolution of 1 $\mu$s and 256 energy bins. They were extracted in the 2--10 keV energy range from all active PCUs (in a given observation) and all layers, and binned into light curves of 10-ms resolution.

\section{X-ray Data analysis and Results}\label{results}

We inspected all observations for the presence of bursts by a careful examination of the light curves binned with different time resolutions. A few were found (besides those reported in \citealt{gogus10short}, we found another short and weak burst in the \rxte\ data of 2010 May 18). In the analyses that follow, we removed the bursts from the event lists by applying intensity filters.
\swift\ and \rxte\ data from a few contiguous observations carried out with the same instrumental setup were combined in order to achieve better statistics and higher signal to noise ratio. For the timing analysis, the data were corrected to the barycentre of the Solar system using the  \cxo/UKIRT position reported in \citet{gogus10short}. 

\subsection{Spatial analysis}\label{diffuse}

\begin{figure}
\resizebox{\hsize}{!}{\includegraphics[angle=0]{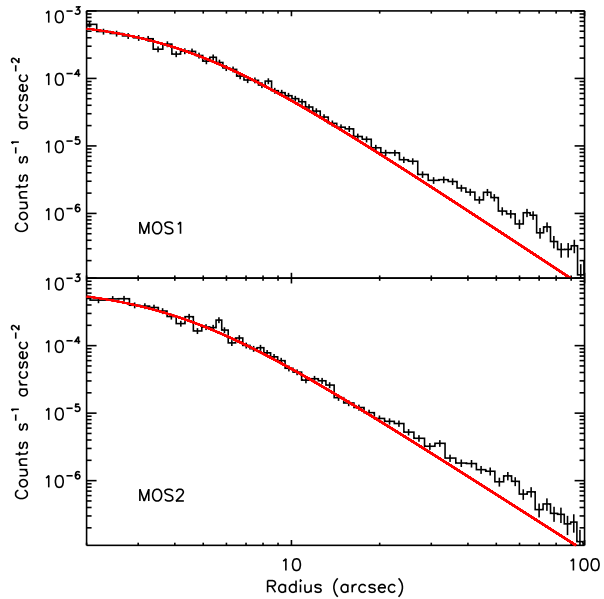}}
\caption{\label{spatial} Surface brightness radial profiles of the background-subtracted X-ray emission of \src\ from the \xmm\ MOS1 and MOS2 cameras (3--6 keV data from the three observations combined). The red solid line shows for each camera the best-fit point-spread function \citep{ghizzardi01}. The radial profiles show clear evidence for extended emission starting from $\sim$35 arcsec.}
\end{figure}

As can be seen in Fig.~\ref{spatial}, some diffuse emission around \src\ was detected during the three \xmm\ observations, mainly in the 3--6 keV energy band. Considering the large absorption derived from the X-ray spectrum, this is likely a halo produced by the scattering of the point source X-ray radiation by dust along the line of sight \citep{overbeck65,tiengo10,rivera10}. 

By analysing the spectrum from different annular regions, we have found that, as expected from a dust halo, the diffuse X-ray emission is significantly softer than the point source. Moreover, the \xmm\ observation of the \src\ field performed in 2006, where no point-like nor diffuse emission was detected at the SGR position (see Section~\ref{archival}), favours the dust halo hypothesis with respect to steady diffuse emission, as expected, for example, in a supernova remnant. 
However, the data quality of the outburst observations is not good enough to detect significant variability in the extended emission at the level expected for a dust scattering halo.

\subsection{Spectral analysis}\label{xspectroscopy}

For the spectroscopy (performed with the \textsc{xspec} 12.6 fitting package; \citealt{arnaud96}) we concentrate first on the \xmm\ spectra which, owing to the EPIC instrument high throughput and long observing time, are those with the best statistical quality. We fit the spectra from the three observations simultaneously with the hydrogen column density tied between all data sets. Photons having energies below 2 keV and above 10 keV were ignored, owing to the very few counts from \src. This resulted in about $6100\pm80$, $5800\pm90$, and $4000\pm70$ net EPIC-pn counts in the three observations in chronological order. The abundances used are those of \citet{anders89} and photoelectric absorption cross-sections are from \citet{balucinska92}.\\
\indent Using the first twelve \swift/XRT observations in PC mode, \citet{gogus10short} could not discriminate between a power-law and a blackbody as the model that better describes the spectrum of the SGR persistent emission. A fit of the \xmm\ pn and MOS data with an absorbed power law yields a relatively high $\chi^2$ value [$\chi^2_\nu=1.21$ for 673 degrees of freedom (dof)] and structured residuals, while a better fit ($\chi^2_\nu=1.05$ for 673 dof) is obtained using a blackbody model (corrected for the absorption). No additional spectral components are statistically required. The results of this simultaneous modelling are presented in Table~\ref{fits} (the values of the spectral parameters were not significantly different when each observation was fit separately). The best-fitting hydrogen column density is $(10.4\pm0.2)\times10^{22}$ cm$^{-2}$ (here and in the following uncertainties are at 1$\sigma$ confidence level, unless otherwise noted) and the blackbody temperature is consistent within the errors with being constant at $kT\simeq1.2$ keV, while the observed flux slightly (but significantly, see Table~\ref{fits}) decreased from $F\simeq3.9\times10^{-12}$ \flux\ to $3.1\times10^{-12}$ \flux\ (in the 2--10 keV band) during the $\sim$20 days that separate the first and last \xmm\ pointings. Assuming an arbitrary distance of 10 kpc, the corresponding \xmm\ maximum and minimum luminosities are $\sim$$9\times10^{34}$ \lum\ and $\sim$$7\times10^{34}$ \lum\ (Table~\ref{fits}). For each \xmm\ observation we performed phase-resolved spectroscopy by extracting the spectra for different selections of phase intervals. No significant variations with phase were detected, all the spectra being consistent with the model and parameters of the phase-averaged spectrum, simply re-scaled in normalisation.

\begin{table*}
\centering
\begin{minipage}{10.5cm}
\caption{Spectral fit of the \xmm\ data with the blackbody model ($\chi^2_\nu=1.05$ for 673 dof). Errors are at a 1$\sigma$ confidence level for a single parameter of interest. The measured \nh\ value is $(10.4\pm0.2)\times10^{22}$ cm$^{-2}$.} \label{fits}
\begin{tabular}{@{}lcccc}
\hline
Obs.ID & $kT$ & Radius$^{a}$ & Absorbed flux$^{b}$ & Luminosity$^{c}$\\
 & (keV) & (km) & ($10^{-12}$ \flux) &($10^{34}$ \lum)\\
\hline
0605851901 & $1.20\pm0.01$ & $0.62\pm0.02$ & $3.9\pm0.1$& $8.9^{+0.2}_{-0.1}$\\
0605852001 & $1.18^{+0.02}_{-0.01}$ & $0.62\pm0.02$ & $3.6\pm0.1$& $8.4\pm0.1$ \\
0605852101 & $1.20^{+0.02}_{-0.01}$ & $0.55\pm0.02$ & $3.1\pm0.1$& $7.2\pm0.1$ \\
\hline
\end{tabular}
\begin{list}{}{}
\item[$^{a}$] The blackbody radius is calculated at infinity and for an arbitrary distance of 10 kpc.
\item[$^{b}$] In the 2--10 keV energy range.
\item[$^{c}$] In the 2--10 keV energy range and for an arbitrary distance of 10 kpc.
\end{list}
\end{minipage}
\end{table*}

Most magnetars, especially when in outburst, exhibit more complex spectra, usually fit by the superposition of two/three blackbodies or a blackbody and a high-energy power law (e.g. \citealt{mte05short,esposito09short,bernardini09short,bernardini11short,icd07,enoto09short,rea09short}). The spectrum of \src\ is reminiscent of that reported  by \citet{esposito10short} for SGR\,0418+5729 during its 2009 outburst. However, at the time of the analysis reported in \citet{esposito10short}, only low-counts-statistics \swift\ and \rxte\ spectra were available for SGR\,0418+5729. Higher quality \xmm\ data, now public, clearly show that a more complex model is required for the emission of  SGR\,0418+5729.\footnote{We have analysed an \xmm\ observation of SGR\,0418+5729 performed on 2009 August 12--13, about 68 days after the outburst onset (obs. ID 0610000601, exposure 67.2 ks). The data are well described by a power law plus blackbody model ($\chi^2_\nu=1.1$ for 231 dof) with the following spectral parameters: $N_{\rm H}=(6.4\pm0.3)\times10^{21}$ cm$^{-2}$, photon index $\Gamma=2.6^{+0.2}_{-0.1}$, blackbody temperature $kT=0.93\pm0.01$ keV, and 0.5--10 keV absorbed flux $(7.0\pm0.2)\times10^{-12}$ \flux. A double blackbody gave a poorer fit ($\chi^2_\nu=1.2$ for 231 dof), whereas both blackbody and power-law models yield statistically unacceptable fits ($\chi^2_\nu\gg2$).}

\begin{figure}
\resizebox{\hsize}{!}{\includegraphics[angle=-90]{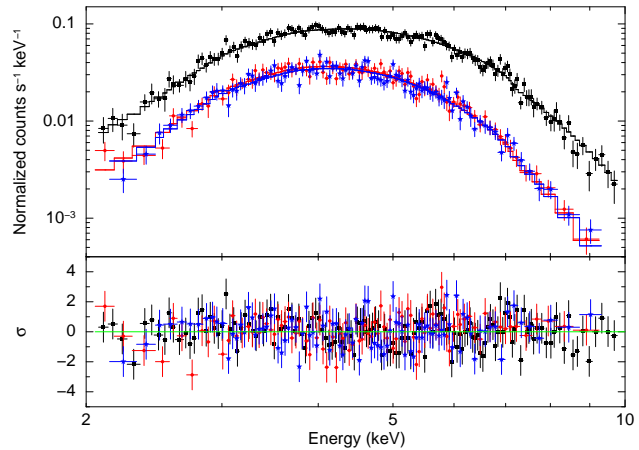}}
\caption{\label{spec} Fit of the spectrum obtained from the first two \xmm\ observations (see Section~\ref{xspectroscopy}) with the single-blackbody model. Black squares, red circles, and blue stars represent the pn, MOS1, and MOS2 data, respectively. Bottom panel: the residuals of the fit (in units of standard deviations).}
\end{figure}

In order to achieve better statistics and higher signal to noise, we merged the data from the first two \xmm\ observations (since there is no evidence for variations in the flux and spectrum of the source between them) and accumulated a combined spectrum which is presented in Fig.~\ref{spec}. Again, the single-blackbody model provides an excellent fit  ($\chi^2_\nu=1.00$ for 352 dof), and no additional components are required. By including in the model a power-law component with photon index fixed to 3 (see e.g. \citealt{rea09short}), we can set $3\sigma$ upper limits of $\approx$25\% and $\approx$30\% on the contribution of this component to the total observed and unabsorbed fluxes, respectively. This shows that the relatively low flux and high absorption of \src\ make even our deep \xmm\ observations not very sensitive to the presence of a second spectral component in this SGR.
We repeated the phase-resolved spectral analysis on the combined dataset. Once more, we see no evidence of spectral 
shape evolution with the rotational phase; in particular, no variation larger than  the 1$\sigma$ error ($k\delta T\simeq0.02$ keV) was found in the blackbody temperatures at different phases.

\begin{figure}
\resizebox{\hsize}{!}{\includegraphics[angle=-90]{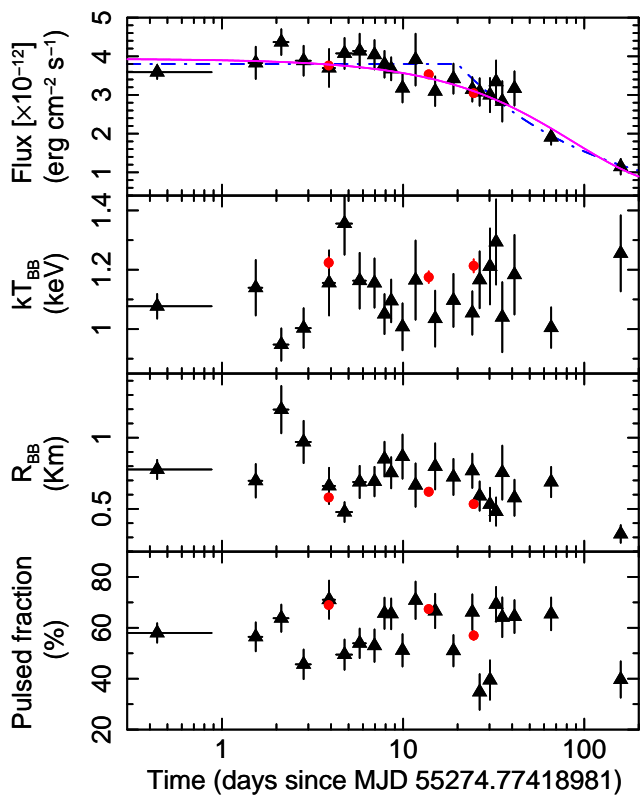}}
\caption{\label{history} Time-evolution of the characteristic parameters of \src\ inferred from the spectral analysis (single blackbody model) of the \swift\ (black triangles) and \xmm\ (red circles) data. The different panels show (from top to bottom): the (absorbed) flux in the 2--10 keV energy range (two possible models describing the decay are also plotted: an exponential function with the solid magenta line and a broken power-law with the dot-dashed blue line; see Section~\ref{xspectroscopy}), the blackbody temperature and radius (evaluated at infinity and assuming an arbitrary distance of 10 kpc), and the pulsed fraction. We assumed as $t=0$ the time of the \swift/BAT trigger.}
\end{figure}

To obtain flux measurements over the outburst, the \swift/XRT data were fit in the same way as the \xmm\ ones, simultaneously and with all parameters left free to vary, except for the absorption column density, that this time was fixed at the value measured with \xmm. This resulted in an acceptable fit ($\chi^2_\nu=1.01$ for 602 dof) with spectral parameters similar to those reported in Table~\ref{fits} (see Fig.~\ref{history}).  We plot the resulting long-term light curve in the top panel of Fig.~\ref{history}. 

Although the available data do not allow us to perform an accurate modelling of the decay shape, because of the relatively high uncertainties on the fluxes and moderate time-span, we observe that similarly good fits can be obtained with either an exponential function or a broken power-law model. For an exponential function of the form $F(t) = A \exp(-t/\tau)$, the best-fitting values ($\chi^2_\nu=0.84$ for 23 dof) are $A=(3.92\pm0.05)\times10^{-12}$ \flux\ and $\tau=(108\pm9)$ day.\footnote{An additional component $F_0$ representing the base flux level [$F(t) = F_0+A \exp(-t/\tau)$] is not required. Including it anyway in the fit  ($\chi^2_\nu\simeq0.83$ for 22 dof), we find $F_0=(5\pm4)\times10^{-13}$ \flux.} Adopting a broken power law ($\chi^2_\nu=0.73$ for 22 dof;), the break occurs at $(20\pm3)$ d, when the index changes from $\alpha_1=-0.01\pm0.02$ to $\alpha_2=-0.54\pm0.09$; the flux at the break time is $(3.7\pm0.2)\times10^{-12}$ \flux.  In both cases we assumed  as $t=0$ the time of the \swift/BAT trigger. From the broken power-law fit in particular, it is apparent that the source flux is consistent with a constant value for the first $\sim$$20$ days. The two models considered are plotted in Fig.~\ref{history}.

\subsection{Timing analysis}\label{xtiming}
  
\indent In \citet{eis10} we reported on our search for periodicities made on the first \swift/XRT observation (00416485000) by calculating a fast-Fourier-transform power spectrum. A very prominent peak occurs in the spectrum of that observation at 7.5653(4) s (the quoted uncertainty indicates the Fourier period resolution). Pulsations were clearly detected also in all the others \swift\ and \xmm\ datasets, and in the \rxte\ ones up to 2010 October 30 (MJD 55499), when presumably the flux became too low for the PCA sensitivity. 

In order to obtain a refined ephemeris for the longest possible baseline, we studied the pulse phase evolution in these observations by means of an iterative phase-fitting technique (see e.g. \citealt{dallosso03}). The fits were carried out in the range 2--10 keV with a $\chi^2$ minimisation approach using \textsc{minuit} \citep{james75}.  

\begin{figure}
\resizebox{\hsize}{!}{\includegraphics[angle=-90]{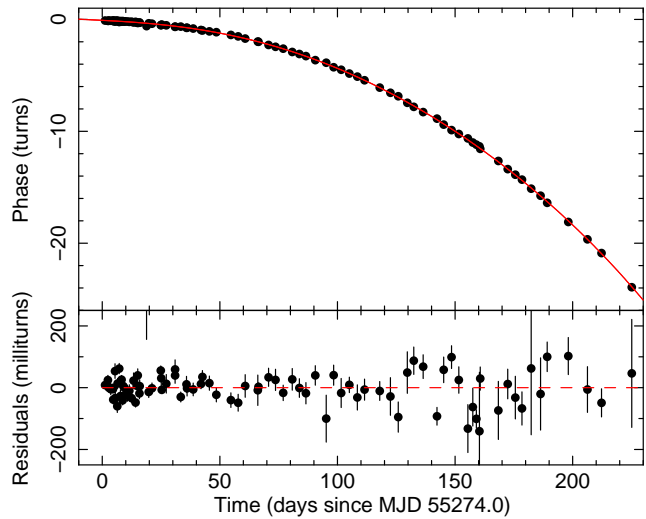}}
\caption{\label{residuals} Top panel: \swift, \xmm,\ and \rxte\ pulse phase evolution  with time with respect to the period measured during the first \swift\ observation. The solid red line represents the best-fitting fourth-order polynomial solution. Bottom panel: Time residuals with respect to timing solution.}
\end{figure}

Throughout the period covered by useful observations ($\sim$225 days) the relative phases and amplitudes were such that the phase evolution of the signal could be followed unambiguously. A second-order polynomial, as employed in the recent analysis by \citet{gogus10short} over the first $\sim$47 days, provides an unacceptable fit to the data, with $\chi^2_\nu=9.06$ for 86 dof. We tried higher-order polynomials until the addition of a further (higher-order) term was not statistically significant at more than 3$\sigma$ with respect to the null hypothesis (as evaluated by the Fisher test). The outcome of this process was a fourth-order polynomial (the improvement obtained in the fit with a fifth-order polynomial has a statistical significance of only 2.6$\sigma$), which we used to fit the phase shifts.  The resulting phase-coherent solution is given in Table\,\ref{timing-fit} and plotted in Fig.~\ref{residuals}; the best fit ($\chi^2_\nu=1.07$ for 84 dof) gives $\nu=0.132\,180\,571(7)$ Hz and $\dot{\nu}=-6.0(5)\times10^{-14}$ Hz s$^{-1}$,  assuming MJD 55274.0 as reference epoch. We have checked that the positional uncertainty ($0\farcs3$; \citealt{gogus10short}) does not significantly affect the rotational parameters resulting from our analysis.

\begin{table}
\centering
\caption{Spin ephemeris of \src\ obtained using the combined \rxte, \swift, and \xmm\ observations. We also give for convenience the corresponding period $P$ and period derivative $\dot{P}$, as well as the derived characteristic age $\tau_c=P/(2\dot{P})$, dipolar magnetic field $B\approx(3c^3IP\dot P/8\pi^2R^6)^{1/2}$, and rotational energy loss $\dot{E}=4\pi^2 I \dot{P}P^{-3}$ (here we took $R=10$ km and $I=10^{45}\ {\rm g\, cm}^2$ for the star radius and moment of inertia, respectively).}
\label{timing-fit}
\begin{tabular}{@{}lc}
\hline
Parameter & \hspace{3cm} Value\\
\hline
 Range (MJD) & \hspace{3cm} 55274.775--55499.170\\
 Epoch (MJD) & \hspace{3cm} 55274.0\\
$\nu$ (Hz) & \hspace{3cm} 0.132\,180\,571(7) \\
$\dot{\nu}$ (Hz s$^{-1}$) & \hspace{3cm} $-6.0(5)\times10^{-14}$\\
$\ddot{\nu}$ (Hz s$^{-2}$) & \hspace{3cm} $-1.3(2)\times10^{-20}$ \\
$\dddot{\nu}$ (Hz s$^{-3}$) & \hspace{3cm} $9(2)\times10^{-28}$ \\
$\chi^2$/dof & \hspace{3cm} 89.47/84\\
\hline
$P$ (s) & \hspace{3cm} 7.565\,408\,4(4) \\
$\dot{P}$ (s s$^{-1}$) & \hspace{3cm} $3.5(3)\times10^{-12}$\\
\hline
$\tau_c$ (kyr) & \hspace{3cm} 35 \\
$B$ (G) & \hspace{3cm} $1.6\times10^{14}$\\
$\dot{E}$ (\lum) & \hspace{3cm} $3.2\times10^{32}$\\
\hline
\end{tabular}
\end{table}

\begin{figure}
\resizebox{\hsize}{!}{\includegraphics[angle=-90]{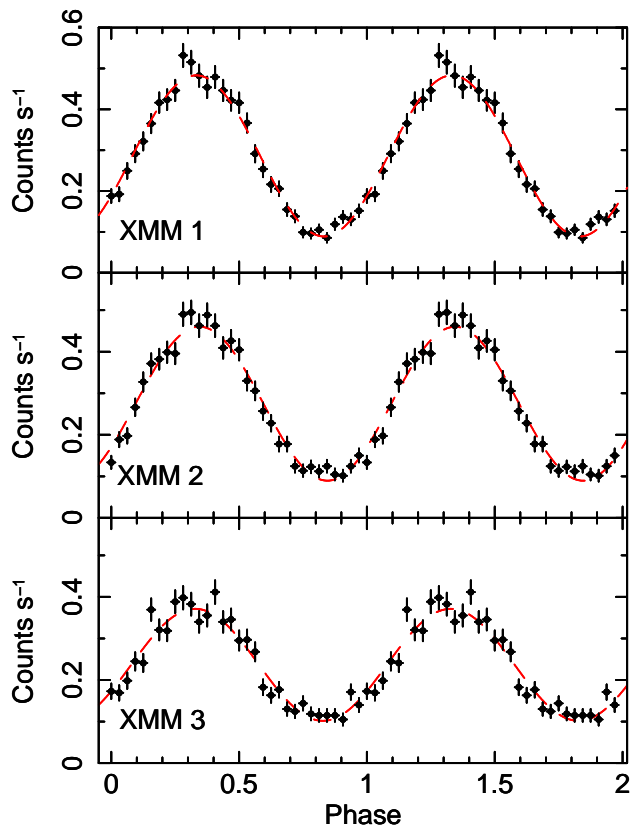}}
\caption{\label{pprofile} Background-subtracted \xmm\ epoch-folded pulse profiles (32-bin, EPIC-pn). The sinusoidal fit to the data is superimposed in red (the $\chi^2_{\nu}$ values are 0.9, 1.0, and 1.2 for 29 dof, for the three observations in order).}
\end{figure}

In Fig.~\ref{pprofile} we show the three \xmm\ light curves obtained folding  the high time-resolution (nominal frame time of 73.4 ms) EPIC-pn data at our phase-coherent ephemeris. The pulse profile is sinusoidal and the pulsed fraction, that we define as the semi-amplitude of sinusoidal modulation divided by the mean source count rate, was consistent through the first two observations [($69.0\pm1.4$)\% and ($67.4\pm1.5$)\%, respectively], while it decreased to ($57.0\pm1.7$)\% in the third one. We have investigated the morphology of the pulse phase distribution as a function of energy by comparing the \xmm\ pulse profiles in different energy bands. The measured pulsed fractions  in the three observations are  $(66\pm2)\%$,  $(64\pm2)\%$, and $(56\pm2)\%$ in the soft (2--5 keV) energy band, and $(75\pm2)\%$, $(71\pm2)\%$, and $(59\pm3)\%$ in the hard (5--10 keV) band. Apart from this marginal indication for an increasing trend of the pulsed fraction with energy, no significant pulse shape variations (such as phase shifts of the maxima) were found as a function of energy by cross-correlating or comparing through a two-sided Kolmogorov--Smirnov test the soft and hard folded profiles.

The 2--10 keV pulsed fractions measured in the individual observations obtained with \xmm\ and \swift\ are shown in the bottom panel of Fig.~\ref{history} (we did not consider the \rxte\ data as the non-imaging PCA instrument does not ensure reliable background subtraction).  A constant fit of the pulsed fraction values derived with XRT\footnote{We did not attempt to fit simultaneously the \xmm\ and \swift\ data in order to avoid possible effects due to the different responses and energy dependence of the effective areas of the of the EPIC and XRT detectors.}  (and shown in Fig.~\ref{history}) does not adequately describe the data ($\chi^2_\nu=4.52$ for 24 dof); however no particular trend is apparent in the pulsed fraction evolution and the simple functions we tried do not yield significantly better fits.

\subsection{Archival searches}\label{archival}

No X-ray source was detected at the \cxo\ position of \src\ \citep{gogus10short} in the \xmm\ observation performed in 2006. In a 20-arcsec circle centred at the SGR coordinates we detected in the 2--10 keV energy band 24 EPIC/pn counts (we did not use the MOS data since the source position fell near CCD gaps in both cameras). The net exposure time was  9.2 ks. Considering the expected background counts (estimated from large surrounding source-free regions) and assuming Poissonian fluctuations, we set a 3$\sigma$ upper limit on the 2--10 keV count-rate of \src\ of $2.3\times10^{-3}$ counts s$^{-1}$. Assuming a blackbody spectrum with $kT= 1.2$ keV and an absorption of $N_{\mathrm{H}}=10.4\times10^{22}$ cm$^{-2}$, this translates into an upper limit on the 2--10 keV observed flux of $\sim$$4\times10^{-14}$ \flux\ and of $\sim$$8\times10^{-14}$ \flux\ on the 2--10 keV unabsorbed flux. If instead we assume a much softer blackbody spectrum with $kT=0.3$ keV, as it might be expected from a magnetar in quiescence (e.g. \citealt{lamb02,gotthelf04,bernardini09short}), the following upper limits are derived (2--10 keV band): $\sim$$2\times10^{-14}$ \flux\ on the observed and $\sim$$2\times10^{-13}$ \flux\ on the unabsorbed flux.

More recently, on 2009 February 13, the field of \src\ was serendipitously imaged for 8 ks by the \cxo/ACIS-S (see also \citealt*{misanovic11}). The SGR was not detected and \citet{gogus10short} derived a 2$\sigma$ upper limit on the absorbed flux of $3.4\times10^{-13}$ \flux, assuming a power-law model with $\Gamma\sim3$ and $N_{\mathrm{H}}\sim14\times10^{22}$ cm$^{-2}$ \citep{gv10}.

\section{Radio observations}\label{radio-obs}

\subsection{ATCA}
We observed the \swift/XRT position of \src\ at 20 cm twice, on 2010 March 22 (UT time 21:40--22:20) and March 23 (UT time 00:17--01:32), with a total on-source integration time of 105 minutes. The observing band was 1.1--1.9 GHz, split in 1 MHz channels and with a central frequency 1.52 GHz; however, several channels, affected by radio frequency interferences, were excised from the spectrum. The array was in the hybrid configuration H168 (maximum baseline $\sim$4.5 km).\footnote{See http://www.narrabri.atnf.csiro.au/observing/configs.html for more details.}

Given the location of the source in the Galactic plane, images including the short baselines were too confused by the emission of extended sources to yield useful limits. Continuum images combining all usable frequency channels were then made considering only the long baselines, i.e. those including the antenna CA06. These images were still somewhat confused by other sources in the field and the rms noise near the field centre was twice that outside the $\sim$$11\arcsec\times80\arcsec$ wide beam. In particular, the local rms at the SGR position turned out to be 0.3 mJy, whence the $3\sigma$ flux density limit for an unresolved continuum radio emission from \src\ is 0.9 mJy.

\subsection{Parkes}

The source was observed on 2010 March 25 (MJD 55279.954) for 90 minutes at a centre frequency of 1374 MHz with the central beam of the Parkes 20 cm multibeam (MB) receiver \citep{staveley96short}. The total 288 MHz bandwidth was split into 96 channels, each 3 MHz wide, and the time series was 1 bit sampled every 1 ms. Data have been searched for periodicities around the value obtained from the X-ray observations over a wide range of dispersion measure values (DM $<$ 6000 pc cm$^{-3}$, given the very high \nh\ value derived from X-ray observations). A search for single de-dispersed pulses was also carried out.

No radio pulsation with a period matching (within $\pm$2.5 ms) the X-ray period, nor half of it, has been found. The single pulse search only revealed the presence of a bright known pulsar in the beam (the 85-ms J1833--0827/B1830--08, at DM 411 pc cm$^{-3}$; see Section~\ref{hess}). The upper limit on the pulsed flux density at 20 cm wavelength is of 0.09 mJy for a duty cycle of 10 per cent, calculated for the maximum DM investigated and assuming a minimum signal-to-noise ratio of 9. The latter relatively high value for the adopted minimum signal-to-noise ratio is due to the presence of strong radio interferences at $\sim$7.5 s, that affect the signal folded at the SGR period.

\section{X-ray emission around PSR\,B1830--08}\label{hess}

\begin{figure*}
\centering
\resizebox{!}{7.6cm}{\includegraphics[angle=0]{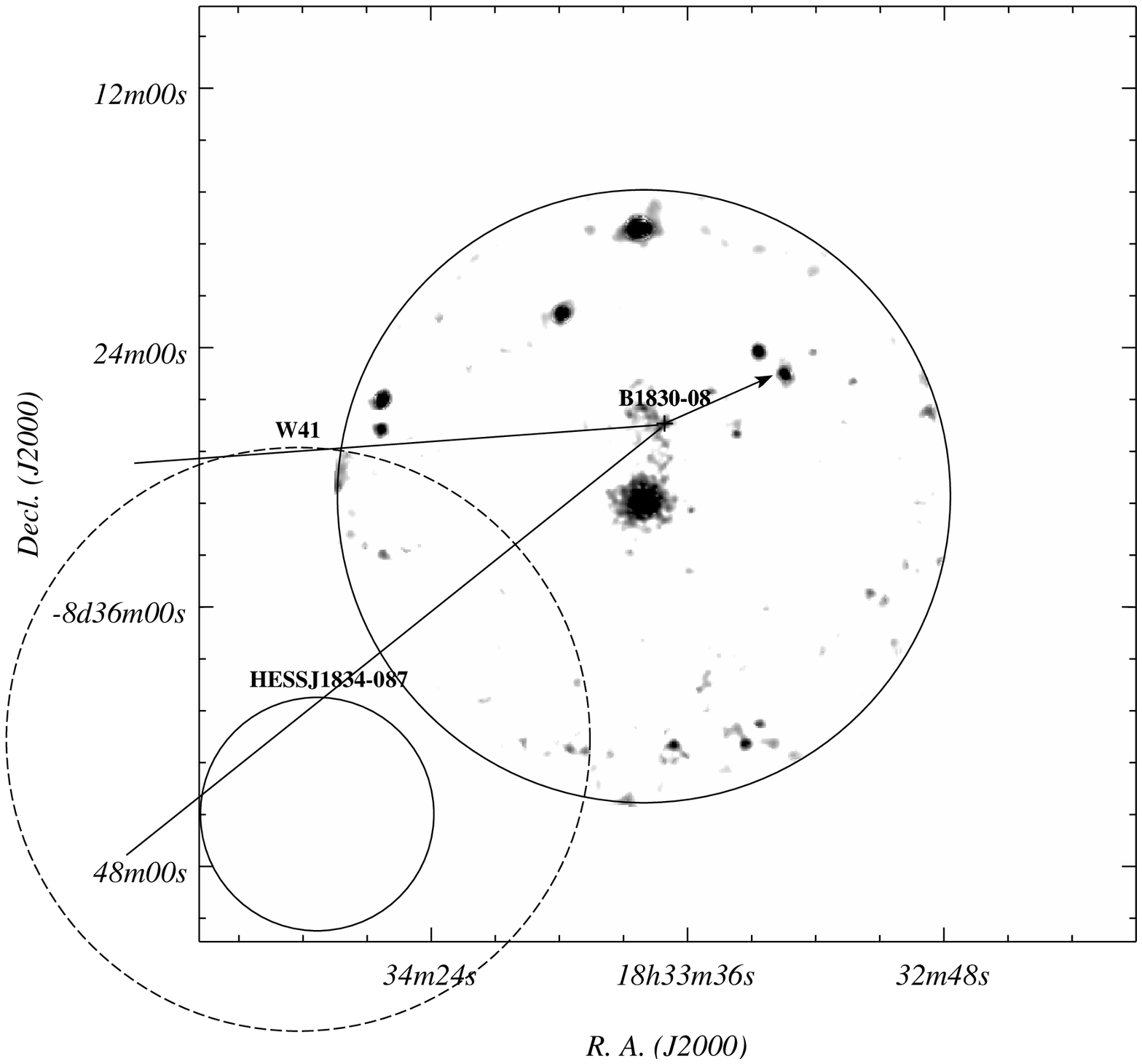}}
\hfill
\setlength\fboxsep{0pt}
\setlength\fboxrule{0.5pt}
\fbox{\resizebox{!}{7.5cm}{\includegraphics[angle=0]{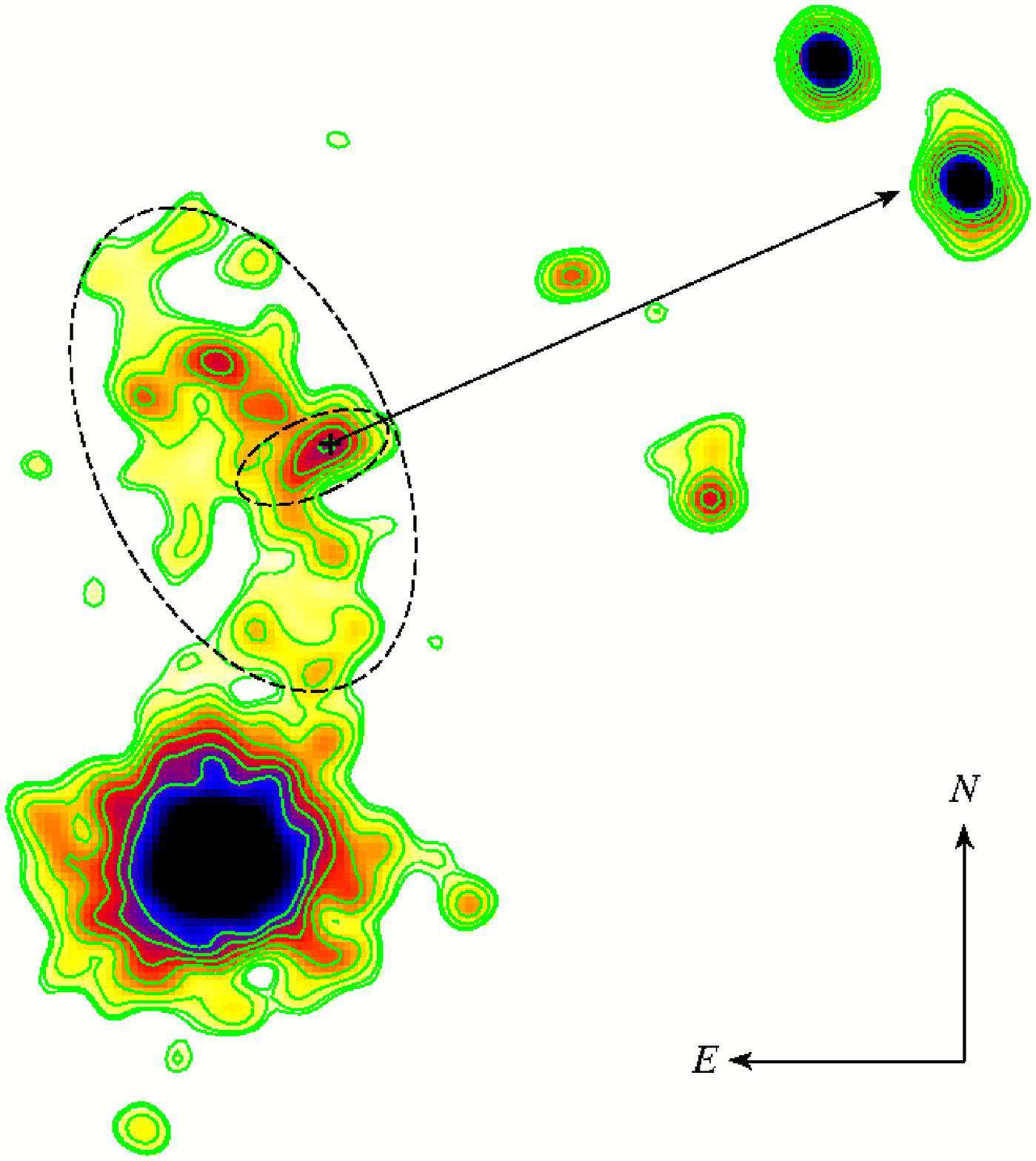}}}
\caption{\label{B1830mosa} {\em Left}: \xmm\ EPIC-MOS mosaiced image. The image is background-subtracted, exposure-corrected and adaptively smoothed; the intensity  scale has been chosen to highlight the faint diffuse emission around PSR\,B1830--08. The brightest source in the centre is \src. The pulsar position is marked with a cross. The arrow indicates the pulsar's measured direction of motion, the length of the arrow corresponding to the distance travelled by the pulsar over 10 kyr. The two solid lines from the pulsar position to east indicate the extrapolation of the pulsar proper motion backwards over the last 50 kyr accounting for the measure uncertainties. The large solid circle encloses the EPIC field of view. The extensions of the supernova remnant W41 \citep[dashed circle;][]{kassim92} and of HESS\,J1834--087 \citep[1$\sigma$, small solid circle;][]{aab06short} are also shown. {\em Right}: zoom-in of the EPIC-MOS mosaiced image shown in the left panel. The contours are uniformly spaced. The inner and outer dashed ellipses outline the `compact' and `halo' emissions, respectively. The length of the compass' arms is 2\arcmin.}
\end{figure*}

Inspection of the \xmm\ EPIC images reveals an excess of X-ray emission north of \src, encompassing the position of the middle-aged radio pulsar PSR\,B1830--08 (spin-down age $\tau_c \sim 150$ kyr; \citealt*{tml93}).

PSR\,B1830--08 is located outside, and possibly associated to, the asymmetric shell-type supernova remnant W41 (G23.3--0.3)  despite an angular separation of 24\arcmin\ to the centre of the remnant \citep{kassim92, gaensler95}. Its distance estimated through the dispersion measure ($\rm DM=411$ pc cm$^{Ð3}$) and the NE2001 model of the Galactic electron density distribution by \citet{cordes02} is 4.6 kpc. PSR\,B1830-08 might be also linked to the unidentified TeV source HESS\,J1834--087 lying inside W41 \citep{aab06short,albert06short}. 
In this case, the extended TeV source would be a relic pulsar wind nebula (PWN) generated by electrons ejected by PSR\,B1830--08 at its birth place and up-scattering the surrounding low-energy photons \citep[e.g.,][]{bartko08}. However, different explanations have been suggested: the TeV emission of HESS\,J1834--087 could result from the $\pi^0$ decay of shock-accelerated hadrons interacting with a giant molecular cloud along the line of sight \citep{tian07} or from the inverse Compton radiation of electrons in a different PWN recently discovered near the centre of W41 (\citealt{mukherjee09}; \citealt*{misanovic11}). The detection of a large-scale X-ray nebula trailing the pulsar in the direction of the supernova remnant and of the TeV source would support the association between PSR\,B1830--08 and HESS\,J1834--087.

We refined the imaging analysis by considering only EPIC-MOS data from the observations 0605852001 and 0605852101, whereas PSR\,B1830--08 was located in the unread regions of the MOS detectors operating in Large Window mode during observation 0605851901. The EPIC-pn data of the three observations were not used either, because of a higher level of residual soft proton contamination indicated by the ratio of the surface brightness inside the field of view to the one outside the field of view \citep{deluca04}. We selected only 1- and 2-pixel events with the default flag mask.

We built background maps using the Filter Wheel Closed (FWC) event lists provided by the \xmm\ background working group.\footnote{See http://xmm.vilspa.esa.es/external/xmm\_sw\_cal/background/index.shtml.} As the profile of the FWC maps changed since the launch of the mission, we filtered only the events collected after January 2007 to produce a more accurate template of the particle-induced background. The background maps were re-scaled in narrow energy bands according to the counts in the non-vignetted MOS corners. Finally, the images were mosaiced, background-subtracted, exposure-corrected, and adaptively smoothed to achieve a signal-to-noise ratio of 6.

Diffuse emission around PSR\,B1830--08 is apparent in the 2--10 keV band mosaic, extending $\sim$2\arcmin\ north-east and south of the pulsar (Fig. \ref{B1830mosa}, left panel). A closer inspection shows a compact structure at the pulsar position embedded in a halo of lower surface brightness (Fig. \ref{B1830mosa}, right panel). The compact structure is $\sim$1\arcmin\ elongated towards south-east, whereas the halo is mostly extending in the perpendicular direction. The bulk of the halo emission lies closer to PSR\,B1830--08, while the the south-western tip is apparently connected with SGR 1833--0832; this is most likely due to the smoothing procedure. Both the halo and the compact structures are hardly visible in the soft band (0.5--2 keV), suggesting an absorbed non-thermal emission. This is confirmed by the spectral analysis.

We produced the spectra also including the three EPIC-pn data sets. The source counts were extracted from the mutually exclusive elliptical regions shown in Fig. \ref{B1830mosa} (right panel). The background counts were estimated from 60\arcsec apertures placed on the same CCD of each camera, and at the same distance to the readout node as the source for the EPIC-pn. For each EPIC camera, we merged the individual spectra from the different observations into a single spectrum and combined the response matrices. The compact and halo regions contained $560 \pm 30$ and $2800 \pm 160$  background-subtracted counts, respectively, corresponding to an average surface brightness of 0.2 and 0.085 counts arcsec$^{-2}$. The spectra were rebinned with a minimum of 20 counts per spectral channel for the compact emission and 50 for the halo.

\begin{figure}
\centering
\resizebox{\columnwidth}{!}{\includegraphics[angle=-90]{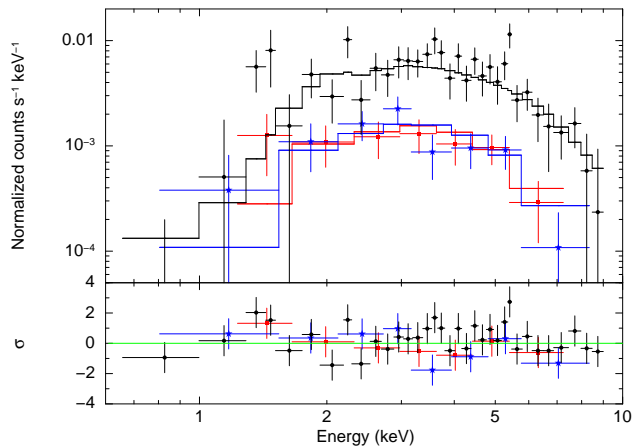}}
\caption{\label{B1830spec} EPIC MOS1 (red circles), MOS2 (blue stars), and pn (black squares) spectra of the compact structure around PSR\,B1830--08 (inner ellipse in Fig. \ref{B1830mosa}, right) fit to the model described in the text. The lower panel shows residuals from the best fit in units of 1$\sigma$.}
\end{figure}

The EPIC pn and MOS spectra of the compact structure (Fig. \ref{B1830spec}) are well fit by an absorbed power-law model, yielding photon index $\Gamma = 1.9^{+0.7}_{-0.6}$, unabsorbed 2--10 keV flux $F_{\mathrm{(2-10\,keV)}} = (1.6^{+0.4}_{-0.2})\times 10^{-13}$ \flux, and hydrogen-equivalent column density \nh$ = (4 \pm 2) \times 10^{22}$ cm$^{-2}$ (reduced $\chi^{2} = 1.0$ for 46 dof). The halo spectra are also well fit by an absorbed power-law model with $\Gamma = 1.7^{+0.5}_{-0.4}$, unabsorbed 2--10 keV flux  $F_{\mathrm{(2-10\,keV)}} = (5.9^{+0.7}_{-0.6}) \times 10^{-13}$ \flux, and  \nh$ = (2.9^{+1.3}_{-0.9})\times 10^{22}$ cm$^{-2}$ (reduced $\chi^{2} = 1.1$ for 153 dof). By fitting simultaneously the compact and halo spectra with the same column density, we found \nh$ = (3.4^{+1.1}_{-0.9})\times 10^{22}$ cm$^{-2}$, slightly higher than the neutral hydrogen column density from radio observations $\sim$$2 \times 10^{22}$ cm$^{-2}$ \citep{dickey90}, which might indicate additional absorbing material along the line of sight.

The hard diffuse emission around PSR\,B1830--08 is suggestive of a PWN. With a spin-down power of $\dot{E}=  5.8\times 10^{35}$ \lum, PSR\,B1830--08 is energetic enough to produce a synchrotron nebula detectable in X-rays for a broad range of distances. For an assumed distance of 4.6 kpc, the measured fluxes imply a luminosity in the 2--10 keV energy range of $\sim$$4 \times 10^{32}$ \lum\ for the compact structure and of $\sim$$1.5 \times 10^{33}$ \lum\ for the halo. This corresponds to a conversion efficiency $L_{\mathrm{X}}/\dot{E}$ of $7 \times 10^{-4}$ and $3 \times 10^{-3}$, within the observed range of pulsar/PWN systems \citep{possenti02, kargaltsev08}.

The inner compact structure looks one-sided, and its brightness peak is coincident with the pulsar position. Such a morphology is reminiscent of a bow-shock PWN, in which the wind of a supersonically moving pulsar is confined by the ram-pressure of the surrounding medium \citep[see][for a review]{gaensler06}. This phase occurs once the pulsar escapes its remnant, several ten thousand years after the supernova event, which is compatible with the age of PSR\,B1830--08. Furthermore, the proper motion of PSR\,B1830--08 ($33\pm5$ mas yr$^{-1}$, corresponding to a linear velocity of $730 d_{4.6}$ km s$^{-1}$; \citealt{hobbs04}) lies in the high-velocity end of the pulsar population and it is most likely supersonic. The compact emission seems to be aligned with the pulsar proper motion (Fig. \ref{B1830mosa}, right), the brightness profile declining gradually from the pulsar position to south-east and sharply to north-west. 

The location opposite to the direction of the pulsar motion, the hard extended emission, and the column density suggest that the halo is also linked to PSR\,B1830--08. However, its morphology is different from observed \citep[e.g., the `Mouse' G359.23--0.82,][]{gaensler04} and simulated \citep[e.g.,][]{bucciantini05} bow-shock PWNe, in which the emission is tightly confined along the proper motion direction. The halo forms instead a broad diffuse region behind the pulsar, almost perpendicular to the proper motion direction.

\section{Discussion}\label{disc}
\subsection{\src}
In this work, we presented a new timing solution for \src\ based on a phase-coherent analysis of \xmm, \swift, and \rxte\ data and valid in the range from MJD 55274.775 to 55499.170. Under the standard assumption that the neutron star slows down because of magnetic braking, it implies a characteristic age $\tau_c\simeq 35$ kyr, a dipolar magnetic field $B\approx1.6\times10^{14}$ G and a rotational energy loss rate $\dot{E}\simeq3.2\times10^{32}$ \lum\ (see Table\,\ref{timing-fit}). 

The modelling of the phase shifts required a fourth-order polynomial which in general is an indication for timing noise (`polynomial whitening'). However, it is worth noting that the second derivative we measured is unlikely to be related to a random change of the pulse profiles, which is expected to introduce only a random distribution of the phase residuals, rather then a cubic term. Recent studies on a sample of 366 radio pulsars  showed  that cubic terms in the phase residuals are actually possible but smaller by several orders of magnitudes than those detected in \src\footnote{The indicator commonly used to quantify the amount of timing noise in radio pulsars is the $\Delta_8$ parameter introduced by \citet{arzoumanian94}. The value computed with the ephemeris of \src\ given in Table\,\ref{timing-fit} is $\Delta_8\simeq4.2$, which is much higher than the typical values measured for radio pulsars (roughly between 0 and $-5$; \citealt{arzoumanian94}; \citealt*{hobbs10}). Moreover, considering the correlation between timing noise and spin-down rate, one would expect for \src\  $\Delta_8$ to be approximately $-0.6$ \citep*{hobbs10}.} and recorded on time-scales longer (years) than those we are sampling in our  dataset \citep*{hobbs10}. We also note that the (long-term) timing noise of young radio pulsars (age $<10^5$yr) can be better understood as resulting from the recovery from previous glitch events \citep*{hobbs10}, making it unlikely that the frequency second derivative is actually due to random noise. 

A possibility is therefore that the higher-order frequency derivatives we observed are manifestations of a glitch recovery. In the AXP 1RXS\,J1708--4009, for instance, negative second frequency derivatives in the $-($0.01$\div$1.3$)\times$10$^{-20}$\,Hz\,s$^{-2}$ range has been detected just after a glitches (\citealt*{dib08}; \citealt{israel07}). We found no evidence for glitches in the time spanned by our observations, but this does not exclude the possibility that such events occurred before the first observation of the outburst.

Another possibility is that the frequency second derivative we measured (or at least a significant fraction of it) is linked to the magnetospheric activity of \src. In the magnetar scenario the spin derivative is in fact expected to increase while the magnetospheric twist is growing, i.e. for instance in periods preceding large outbursts, and to decrease in the aftermath (see e.g. \citealt*{tlk02,beloborodov09}). On the other hand, as pointed out by  \citet{beloborodov09}, a negative $\ddot{\nu}$ following a period of bursting activity (as that detected here) can be still accounted for if a magnetospheric twist is suddenly implanted but its strength is moderate (twist angle less than $\sim$1 rad). Then the twist may still grow for a while in spite of the luminosity released by dissipation monotonically decreases. Only after the twist angle has reached its maximum value, it will start to decay, together with the torque.

Nevertheless, the timing properties of the source appear consistent with those typically observed in the AXP/SGR class. The spectral characteristics of \src\ are instead somewhat less usual. The X-ray spectrum is well described by a single blackbody with temperature $kT\sim1.2$ keV, definitely higher than what generally observed both in transient and `persistent' magnetars ($kT\la 0.7$ keV, with the exception of SGR\,0418+5729; \citealt{esposito10short,rea10short}). We cannot however exclude that the high temperature of \src\ partially results from a bias in the spectral fits that, given the paucity of counts below $\sim$2--3 keV (owing to the large absorption), acts in the direction of increasing the temperature, in order to account for a second spectral component which cannot be properly modelled (see Section~\ref{xspectroscopy}).

The large increase in flux, a factor $\ga$20 above the quiescent level, is indicative of a rather powerful event, not dissimilar from those observed in other transient sources. This, however, seems to be in contrast with the low level of activity seen in \src, from which only few bursts were detected \citep{gogus10short}. A scenario in which we are presently witnessing the later stages of an outburst which was caught during its decay rather than at its onset, could account for the low activity but seems difficult to reconcile with the current high value of the temperature (if the thermal component in the spectrum of \src\ is indeed at $kT\sim 1.2$ keV). A further possibility might be that \src\ experienced a (crustal) heating episode which, however, was not accompanied by (or did not trigger) a twisting of the external field, as possibly suggested by the non-detection of a hard X-ray tail. In this case, however, an alternative explanation, not related to the twist evolution, for the presence of a negative $\ddot{\nu}$ must be sought (perhaps a long-term postglitch recovery, see above).

Immediately after the BAT trigger that led to the discovery of \src, its observed flux was of $\sim$$3.8\times10^{-12}$ \flux, and it decreased  by $\approx$75\% during the following 5 months (which is not unusual for magnetars; e.g. \citealt{rea11}), to $\sim$$1.1\times10^{-12}$ \flux. Although present data do not allow us to discriminate among different decay patterns, the flux evolution can be satisfactorily described by a broken power law, similarly to the case of SGR\,0418+5729. The flux of \src\ remained fairly constant for about 20 days ($\alpha_1\sim0$), when the decay index changed to $\alpha_2\sim-0.5$. Interestingly, also the decay of SGR\,0418+5729 exhibited a break at $\sim$20 days \citep{esposito10short}.
As in the case of SGR\,0418+5729 (see the discussion in \citealt{esposito10short}), the presence of a break, together with the rather flat decay indices, might be difficult to reconcile with the predictions of the deep crustal heating scenario \citep*{let02}, $F\propto t^{-n/3}$ with $n\sim 2$--3. This model provides a satisfactory description of the flux decay in SGR\,1627--41 (\citealt{kouveliotou03short}, although subsequent analyses found no evidence for the plateau around $\sim 400$--800 d predicted by the model; \citealt{mereghetti06}) and AXP 1E\,2259+586 \citep{zhu08}. At the same time, however, it cannot explain the variety of decay patterns which emerged from recent observations of SGRs/AXPs (e.g. \citealt{rea09short}). This may reflect the intrinsic limitations of the model (which relies an a simplified, one-dimensional treatment), or point towards the existence of different heating mechanisms which are at work during the outbursts of magnetars, like in the case of \src.

Prompted by previous detections of a transient (pulsed) radio emission following X-ray transient activity in other magnetars  \citep{camilo06,camilo07,burgay09}, we observed \src\ with the ATCA and the 64-m Parkes radio telescope \citep{burgay10short,wieringa10short}. No evidence for  radio emission was found, down to flux densities of 0.9 mJy and 0.09 mJy for the continuum and pulsed emissions, respectively. Similar upper limits on the pulsed radio emission, 0.1 mJy at 1.38 GHz and 0.2 mJy at 2.28 GHz  for a 10\% duty cycle, were obtained with the Westerbork Radio Synthesis Telescope in the same days (but not simultaneously with our observations; \citealt{gogus10short}). Despite this rather intensive coverage, the negative results of radio searches performed so far on \src\ cannot anyway be taken as conclusive  because of the rapid variability of the pulsed flux shown by the known radio magnetars (see e.g. \citealt{burgay09}). Moreover, for a distance $d=10$ kpc these limits translates into a pseudo-luminosity $L=Sd^2\approx 10$ mJy kpc$^2$, which is significantly smaller than the 1.4 GHz luminosity of the other known radio magnetars at their peak ($\sim$100--400 mJy kpc$^2$; \citealt{camilo06,camilo07,levin10short}) but still much larger than the luminosities of some known ordinary radio pulsars.\footnote{See the online version of the Australia Telescope National Facility (ATNF) pulsar catalogue \citep{manchester05} at http://www.atnf.csiro.au/research/pulsar/psrcat/.} 

\subsection{PSR\,B1830--08 and HESS\,J1834--08}

The discovery of the X-ray PWN generated by PSR\,B1830--08, suggestive of a pulsar bow-shock and a possible diffuse emission trailing the pulsar proper motion, provides new elements to the identification of HESS\,J1834--087. Accounting for a systematic 20\% uncertainty, the distance of PSR\,B1830--08 is compatible with the one of W41,  $4.0\pm0.2$ kpc \citep{tian07}, which is likely associated to HESS\,J1834--087 given the low probability of a chance superposition. The pulsar spin-down age is also consistent with the dynamical age of remnant, roughly 100 kyr \citep{tian07}. As pointed out by \citet{mukherjee09}, the pulsar moves fast enough to have reached its current position starting only 40 kyr from the geometric centre of the remnant (see also Fig. \ref{B1830mosa}, left panel).  

However, the Gamma-ray to X-ray flux ratio, $F_{\mathrm{(E\,>\,1\,TeV})}/F_{\mathrm{(2-10\,keV)}} \sim 50$ for the compact emission and $\sim$10 for the total X-ray emission, is lower than the one expected from the inverse correlation between this quantity and $\dot{E}$  \citep{mattana09short}, which should be around 600 for PSR\,B1830--08. If HESS\,J1834--08 is the relic PWN generated by PSR\,B1830--08, it is not very efficient in radiating in TeV gamma rays. Otherwise, it may be unrelated to PSR\,B1830--08. The non-detection of a more extended X-ray emission trailing the pulsar does not support the association with HESS\,J1834--087 either. Higher resolution and longer X-ray observations are needed to establish the nature of the halo and the detailed morphology of the compact emission, as well as singling out a possible contribution from the pulsar magnetosphere to the compact emission. 

\section*{Acknowledgments}
We thank the referee, Vicky Kaspi, for valuable and constructive comments. This research is based on observations obtained with \emph{XMM-Newton}, an ESA science mission with instruments and contributions directly funded by ESA Member States and NASA.  We thank Norbert Schartel and the staff of the \xmm\ Science Operation Center for promptly scheduling and executing our observations. The \rxte\ and \swift\ data were obtained through the HEASARC Online Service, provided by the NASA/GSFC. We thank the \swift\ PI, Neil Gehrels, the \swift\ duty scientists and science planners for making our \swift\ target of opportunity observations possible. We thank Hans Krimm for re-analyzing the \swift/BAT transient monitor and survey data for the time period following the outburst. We thank Philip Edwards of CSIRO for prompt allocations of observing time at the Parkes Observatory and the ATCA. The Parkes Observatory and the ATCA are part of the Australia Telescope, which is funded by the Commonwealth of Australia for operation as a National Facility managed by CSIRO. The Italian authors acknowledge the partial support from ASI (ASI/INAF contracts I/009/10/0, I/011/07/0, I/010/06/0, I/088/06/0, and AAE~TH-058). PE acknowledges financial support from the Autonomous Region of Sardinia through a research grant under the program PO Sardegna FSE 2007--2013, L.R. 7/2007 ``Promoting scientific research and innovation technology in Sardinia''. FM and DG acknowledge the CNES for financial funding. NR is supported by a Ram\'on~y~Cajal fellowship.

\bsp

\label{lastpage}

\end{document}